# Halogenation of Imidazolium Ionic Liquids. Thermodynamics Perspective

Vitaly Chaban

Instituto de Ciência e Tecnologia, Universidade Federal de São Paulo, 12231-280, São José dos Campos, SP, Brazil.

**Abstract**. Imidazolium cations are promising for anion exchange membranes, and electrochemical applications and gas capture. They can be chemically modified in many ways including halogenation. Halogenation possibilities of the imidazole ring constitute a particular interest. This work investigates fluorination and chlorination reactions of all symmetrically non-equivalent sites of the imidazolium cation. Halogenation of all carbon atoms is thermodynamically permitted. Out of these, the most favorable site is the first methylene group of the alkyl chain. In turn, the least favorable site is carbon of the imidazole ring. Temperature dependence of enthalpy, entropy, and Gibbs free energy at 1 bar is discussed. The reported results provide an important guidance in functionalization of ionic liquids in search of task-specific compounds.

TOC Graphic

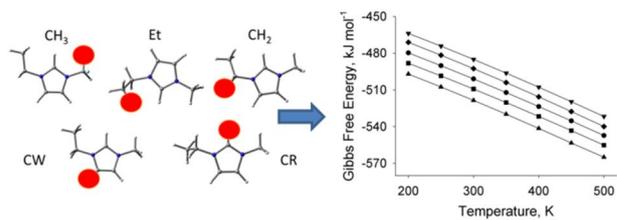

**Introduction**

Imidazolium-based ionic liquids (ILs) remain at the spotlight of chemists and physicists for last years.[1] These ILs are actively investigated for applications in electrochemical devices, gas capture and anionic membranes.[2-7] The imidazolium cations can be coupled with many organic and inorganic anions to provide reasonable tuning of their physical chemical properties.[8-15] Apart from that, the imidazolium cation can be modified chemically. For instance, manifold functional groups can be grafted to the alkyl chains, as well as length of these chains can be tuned. The ethyl, butyl, hexyl, octyl, decyl chains are widely explored around the world.[12,15,16] Addition of a single methylene group results in a significant shear viscosity increase and melting or glassy temperature adjustment. Asymmetric chains of the IL cations play an important role in prohibiting their crystallization, because an energetically favorable crystalline structure becomes impossible.

Halogenation is a well-recognized pathway to modify properties of organic compounds. Many representatives of the family of chlorinated hydrocarbons are omnipresent in chemical laboratories as organic solvents. Fluorination and subsequent polymerization of ethylene resulted in polytetrafluoroethylene, the best known commercial brand of which is Teflon. Carbon fluoride is an extremely stable gas with a low boiling point. It is sometimes used as a low temperature refrigerant. Due to its physical properties, carbon tetrafluoride does not deplete ozone layer and contributes to the greenhouse effect. Halogenation changes interaction sites of the compound. For instance, fluorination or chlorination of the imidazole ring will prohibit hydrogen bonding with the applicable anions and molecular cosolvents. This would have drastic consequences on the derivative ILs, such as different shear viscosity and different phase behavior. Fluorination of the hydrocarbon chains is also interesting, since it allows to obtain even more asymmetric cations and less favorable crystalline arrangements.

Almantariotis and coworkers[17] reported solubility and thermodynamics of solvation of carbon dioxide uptake by the imidazolium-based ionic liquids, 1-octyl-3-methylimidazolium bis[trifluorosulfonyl]imide (TFSI), 1-decyl-3-methylimidazolium TFSI and 1-(3,3,4,4,5,5,6,6,7,7,8,8,8)-tridecafluorooctyl)-3-methylimidazolium TFSI. The properties were determined experimentally at atmospheric pressure and between 298 and 343 K. According to the authors, carbon dioxide solubility is systematically higher in the fluorinated ionic liquid. Carbon dioxide solubility is also higher in the imidazolium IL with a longer hydrocarbon chain. Note, however, that no fluorination of the imidazole ring was performed in this work.

Carvalho and coworkers[18] support the claim that adsorption of sour gases can be adjusted by fluorination of ILs. These authors considered gas-liquid equilibrium of 1-butyl-3-methylimidazolium acetate 1-butyl-3-methylimidazolium trifluoroacetate with carbon dioxide at temperature up to 363 K and external pressures up to 76 MPa. Both anions exhibit a simultaneous interaction of the two oxygen atoms of the carboxylate group with $CO_2$. However, acetate acts as a stronger Lewis base than trifluoroacetate. This was confirmed by ab initio calculations using the reliable electron-correlation post-Hartree-Fock method.

Wang and coworkers[19] synthesized novel hydrophobic ILs having performed partial substitution of chlorine and fluorine in the 1-butyl-3-methylimidazoolium cation, $C_8N_2H_{15}$, via direct chlorine gas treatment and potassium fluoride, respectively. The following two compounds were characterized, $[C_8H_{(12.28)}Cl_{(0.96)}F_{(1.76)}N_2]$ chloride and $[C_8H_{(12.75)}Cl_{(1.58)}F_{(0.67)}N_2]$ hexafluorophosphate. According to these authors, chlorination mainly takes place in the imidazole ring and in the methyl group of the butyl chain. Shear viscosity and hydrophobicity of both halogenated derivatives increased greatly, while the decomposition temperature decreased to certain extent.

This work investigates halogenation (fluorination and chlorination) of the imidazolium-based ILs from the thermodynamics perspective. We identify five chemically non-equivalent prospective halogenation sites in the 1-ethyl-3-methylimidazolium cation. Not only halogenation of side hydrocarbon chains was considered, but also halogenation of the imidazole ring was probed. The substitutional halogenation was shown to be thermodynamically permitted in all cases, whereas fluorination appeared much more energetically favorable than chlorination.

Highly accurate gas-phase ab initio coupled cluster electronic structure calculations were used to obtain molecular partition functions, which can be further processed to obtain thermodynamics quantities, such as enthalpy, entropy, free energy of reaction and their temperature dependences. Demand for thermodynamics data far exceeds current capabilities of the available experimental measurements. Numerical techniques to estimate gas-phase thermodynamics are able to provide certain assistance, that appears of great help for scheduling proper chemical syntheses and interpretation of the already generated experimental data.

**Methodology**

This work reports enthalpy H, entropy S, and Gibbs free energy G for the halogenation reactions involving imidazolium cation. Temperature dependences of each thermodynamics potential are discussed over the range between 200 and 500 K. Note that imidazolium cations are thermally rather stable, whereas thermal decomposition of imidazolium-based ILs normally starts from the anion.

Ab initio calculations do not directly produce enthalpies of formation or any other thermochemical functions. Instead, they produce total molecular energies and optimized electronic structures. The total potential energy is negative and represents changes upon assembling a particle from multiple nuclei and electrons. Entropies, heat capacities, and other quantities are derived from the computed molecular partition function using equations of statistical mechanics. Enthalpy of formation can be derived by computing the energy change for selected chemical reaction. All thermodynamics predictions computed in this way correspond to an ideal gas. The major errors are expected to come from an electronic energy contribution, therefore electron correlation must be described thoroughly to obtain trustworthy results.

An electronic structure of all systems was optimized using the coupled cluster technique. Coupled cluster[20] is a numerical technique, which is used to describe many-body electronic systems and belongs to the group of post-Hartree-Fock methods. The molecular orbitals obtained from conventional one-electron Hartree-Fock calculations are used to construct multi-electron wave functions by means of exponential cluster operator. This allows to explicitly account for electron-electron correlations, which are critically important for thermodynamics potentials derived from an electronic partition function. This method provides highly accurate energies and consequently molecular geometries. However, the underlying computational cost is much higher than that of Hartree-Fock density functional theory computations. Coupled cluster cannot be used for sufficiently large systems, such as periodic supercells. The implementation of coupled cluster employed in the present work uses single and double substitutions from the Hartree-Fock determinant. Furthermore, it includes triple excitations non-iteratively.[21]

The 6-311G Pople-type basis set where polarization and diffuse functions were supplemented to all atoms was applied. No pseudopotentials were applied. That is, all electrons were considered explicitly at all calculation steps. The wave function convergence criterion at every self-consistent field (SCF) step was set to $10^{-8}$ Hartree. No additional techniques to enhance the convergence were applied. The implementation of the outlined electronic structure methods in GAMESS[22] was employed.

The major approximation beyond the reported data is that these thermodynamics potentials assume all non-interacting molecules and ions. Strictly speaking, the results apply to an ideal gas and may, therefore, contain errors depending on the extent that any real condensed-matter system is not ideal. Furthermore, excited states are not considered, which is unlikely principal for imidazolium-based ionic liquids where the band gap is fairly large.

**Results and Discussion**

Figure 1 depicts prospective positions of the halogen substitutions. Positions "$CH_3$", "Et", and "$CH_2$" correspond to alkyl chains, which are grafted to both nitrogen atoms of the imidazole ring. Positions "CR" and "CW" correspond to the carbon atoms within the imidazole ring. All these substitution reactions must be doable according to conventional chemical wisdom. However, it is unclear without a comprehensive numerical analysis, which pathway is more preferable. Especially, the difference between CR and CW is interesting. Despite essentially similar positions in the imidazole ring, the hydrogen atom at CR participates in hydrogen bonding with many anions and polar molecular solvents, whereas the hydrogen atom at CW does not.

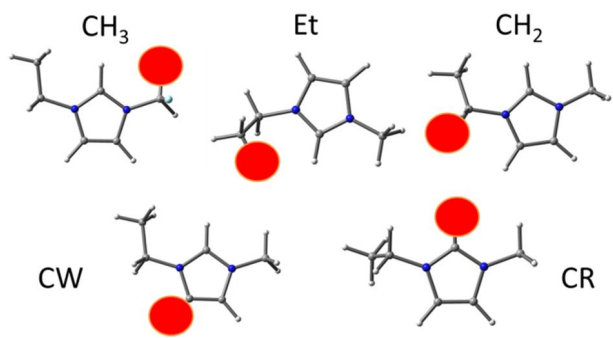

Figure 1. Halogen substitution sites in the imidazolium cation. Carbon atoms are grey, hydrogen atoms are white, nitrogen atoms are blue, and the halogen (fluorine, chlorine) atom is red. Five symmetrically non-equivalent carbon atoms are available. As it will be shown below, these sites are also thermodynamically non-equivalent.

Gibbs free energy change (Figure 2) is an univocal indicator of reaction direction at constant temperature and pressure. While enthalpy change (Figure 3) is almost completely dependent on the strength of chemical interactions, free energy change includes an entropic contribution (Figure 4), which may appear important to help distinguishing between energetically similar processes.

As anticipated, the imidazolium cation can be easily fluorinated at all carbon atoms. At 300 K, the free energy gain due to fluorination ranges from 519 to 485 kJ mol$^{-1}$ depending on the particular substitution site. The process becomes even more energetically favorable as temperature increases. Temperature increase by 300 K (from 200 to 500 K) results in ca. 14% increase of the free energy. The most favorable site is the methylene, $CH_2$, group of the ethyl chain. Next goes the "Et" site, thus, the same alkyl chain is fluorinated. Fluorination of the methyl group is somewhat less favorable, while fluorination of the "CR" and "CW" sites are least favorable. The largest difference between prospective fluorination sites at the same temperature amounts to 9 kJ mol$^{-1}$. Although this difference is fairly modest, it exceeds kT at any of the considered temperatures.

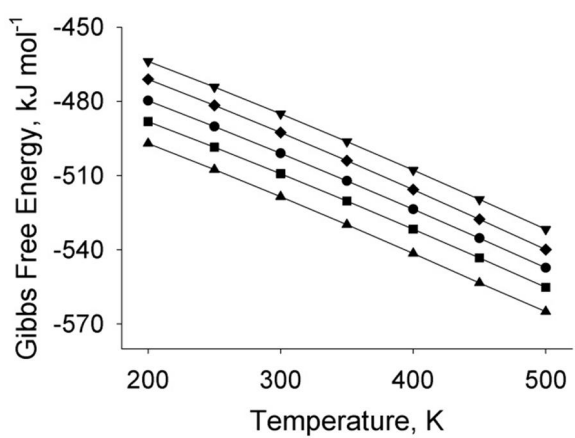

Figure 2. Free energy change upon the fluorination reaction of the imidazolium cation, which takes place via the $CH_3$ (circles), Et (squares), $CH_2$ (triangles up), CW (triangles down), and CR (diamonds) reaction sites.

Enthalpy change (Figure 3) exhibits the same trend, except that the "$CH_3$" and "Et" sites go in the inverse order. Indeed, entropy gain (Figure 4) from the fluorination of "Et" is systematically higher than that from the fluorination of "$CH_3$". Entropy gain increases as temperature increases. Overall, both enthalpy change and entropy change suggest that the considered substitution reaction is thermodynamically permitted.

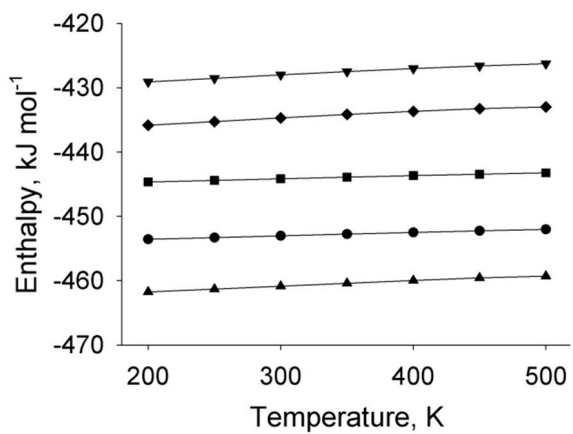

Figure 3. Enthalpy change upon the fluorination reaction of the imidazolium cation, which takes place via the CH$_3$ (circles), Et (squares), CH$_2$ (triangles up), CW (triangles down), and CR (diamonds) reaction sites.

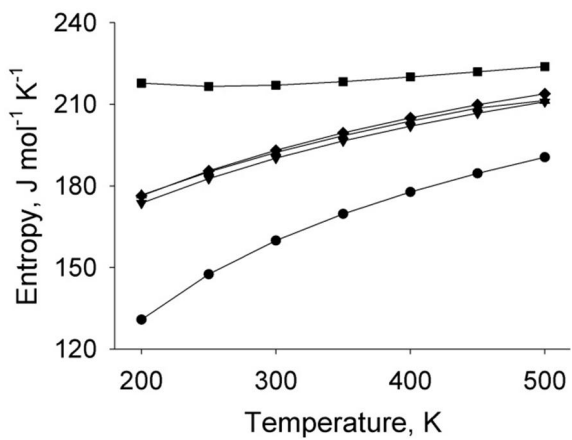

Figure 4. Entropy change upon the fluorination reaction of the imidazolium cation, which takes place via the CH$_3$ (circles), Et (squares), CH$_2$ (triangles up), CW (triangles down), and CR (diamonds) reaction sites.

Chlorination is in many aspects similar to fluorination with a correction for the much higher electronegativity of fluorine and the larger covalent radius of chlorine. Chlorination of aliphatic and aromatic hydrocarbons is omnipresent in organic synthesis. According to Figure 5, chlorination of all substitution sites of the imidazolium cation is also possible. Nevertheless, it is much less energetically favorable, as compared to fluorination. Furthermore, the recorded temperature dependence is different. Temperature increase is unfavorable for the chlorination in the case of the "CH$_3$", "Et", and "CW" sites. Nearly no temperature dependence is observed in the case of the "CH$_2$" and "CR" sites.

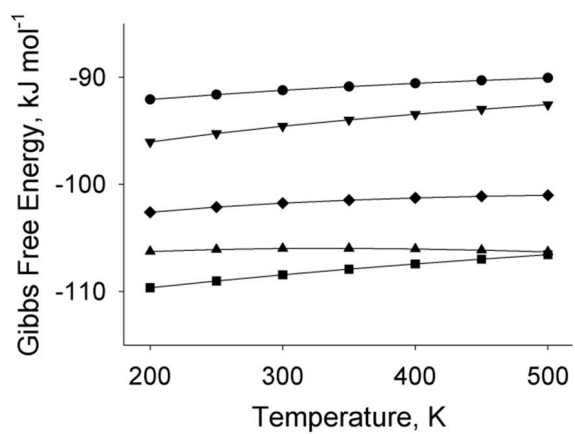

Figure 5. Free energy change upon the chlorination reaction of the imidazolium cation, which takes place via the $CH_3$ (circles), Et (squares), $CH_2$ (triangles up), CW (triangles down), and CR (diamonds) reaction sites.

Specific heat capacity, computed from Figure 6, is notably small for all reactions, which is a difference from fluorination. Recall that chlorination of hydrocarbons is often photochemical, although catalytic chlorination is also well-known. Priority of the chlorination sites is also somewhat different (although not fundamentally) in comparison with fluorination. The "CW" and "CR" sites become more favorable than the "$CH_3$" site. Chemical background of this trend in not immediately evident. Contrary to possible expectations, geometry perturbations of the cation structure are very similar in both the case of fluorination and chlorination. The distances from the halogen atom to the adjacent nitrogen atom of the imidazole ring are 2.67 Å (fluorination) and 2.69 Å (chlorination). The angles F-C-N and Cl-C-N are close to 110 degrees. Further investigation was performed to provide a satisfactory interpretation of the observed trend. The methyl group is responsible for maintaining a positive charge of the imidazole ring. The total electronic charge on $CH_3$ in the thoroughly optimized geometry equals to +0.39e. The halogen atoms obviously somewhat perturb this electron localization shifting an excessive charge to the ring. The sum of charges on $CH_2F$ is +0.37e, whereas the sum of charges on $CH_2Cl$ is +0.18e. Despite a strong electronegativity of fluorine as an element, its effect in the $CH_3$ group is inferior to that of chlorine. The same sort of analysis allows to understand the effect of halogens in the methylene group of another hydrocarbon chain. The sum of non-modified $CH_2$ equals to +0.23e. Compare this value to +0.10e (CHF) and +0.05e (CHCl). The difference between CHF and CHCl is smaller than the difference between $CH_2F$ and $CH_2Cl$.

Note an excellent coincidence of the present results with the experiment by Wang and coworkers.[19] Chlorination of the imidazole ring and the methyl group occurs in the experiment and is thermodynamically most favorable in our simulations. For unclear reasons, chlorination of the methylene groups were not reported by Wang, possibly due to technical difficulties in identification.

In the considered case of chlorination, free energy of reaction occurs mostly due to the enthalpy evolution (Figure 6), rather than due to the entropy evolution (Figure 7). The entropy contribution to the chlorination reaction is generally very modest, which is not the case for fluorination. That is, thermodynamics of the fluorination and chlorination reactions differ significantly.

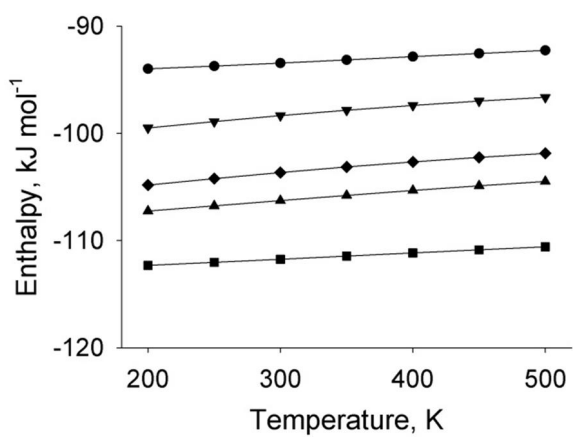

Figure 6. Enthalpy change upon the chlorination reaction of the imidazolium cation, which takes place via the $CH_3$ (circles), Et (squares), $CH_2$ (triangles up), CW (triangles down), and CR (diamonds) reaction sites.

The entropy decreases upon chlorination (Figure 7), except for chlorination of $CH_2$ at high temperatures, 400-500 K. The entropy alteration becomes more positive as temperature of reaction increases, although still remains negative in most investigated cases. Since the entropy change is small, chlorination is determined by the enthalpy factor.

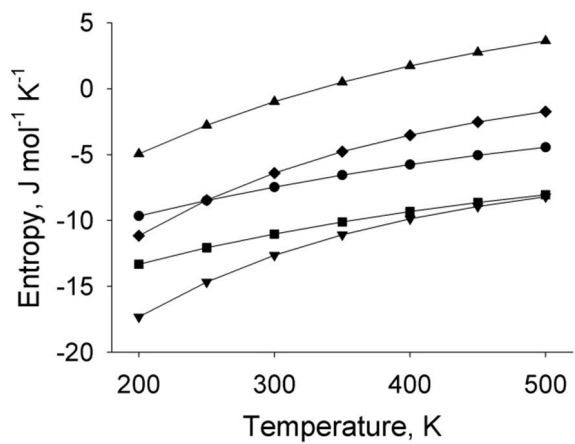

Figure 7. Entropy change upon the chlorination reaction of the imidazolium cation, which takes place via the $CH_3$ (circles), Et (squares), $CH_2$ (triangles up), CW (triangles down), and CR (diamonds) reaction sites.

In an attempt to understand the atomistic background of the observed thermodynamics potentials, Table 1 summarizes partial electron charges of hydrogen, fluorine and chlorine atoms at the equivalent sites of the imidazolium cation. These charges were derived according to the well-known Mulliken scheme. Although being routinely criticized for no physical meaning, Mulliken charges constitute a straightforward and convenient measure of electron distribution within a single molecule. These charges should not be used as absolute values in electron units, but their trends help in understanding of many chemical phenomena, such as those in the present work. Hydrogen carriers a positive charge at any position within the imidazole cation. The most electron deficient hydrogen atom, in the pristine imidazolium cation, +0.23e, is linked to the "CR" site. It is in accordance with hydrogen bonding ability of this cation, which happens particularly through CR-H and does not happen through CW-H. When fluorination occurs at "CR", the partial charge on chlorine becomes -0.10e. In turn, when chlorination occurs, the partial charge becomes +0.24e. Fluorine significantly perturbs an electronic structure of the positive charged ring while chlorine does not. The multiplicity of compound remains singlet in all cases, and the cation has to maintain the +1e charge. The charge analysis provides understanding why the trend exhibited by thermodynamics potentials occurs.

Table 1. Partial electron charges on the investigated reaction sites of the imidazole cation before substitution (hydrogen) and after substitution (fluorine and chlorine). The change of partial charges is helpful in understanding thermodynamics potentials

| Site | Electronic charge, e | | |
|---|---|---|---|
| | Fluorine | Chlorine | Hydrogen |
| $CH_3$ | -0.09 | +0.24 | +0.18 |
| Et | -0.16 | +0.24 | +0.18 |
| $CH_2$ | -0.16 | +0.17 | +0.21 |
| CW | -0.11 | +0.52 | +0.21 |
| CR | -0.10 | +0.24 | +0.23 |

**Conclusions**

Thermodynamics of fluorination and chlorination of the imidazolium cation has been hereby reported. The presented data are based on very accurate ab initio post-Hartree-Fock calculations using the coupled cluster technique. The results show that all possible substitution sites in the imidazolium cation undergo favorable halogenation, including those in the imidazolium ring. The free energy difference between various sites is not drastic, but it exceeds kT at all investigated temperatures. Halogenation of the alkyl chain appears somewhat more favorable than halogenation of the imidazole ring. Thus, the alkyl chain will be halogenated first, unless it is specifically protected beforehand. Fluorination is definitely a more favorable process than chlorination.

Halogenation of the imidazolium-based ionic liquids constitutes a significant interest, since it would allow to adjust their cation-anionic structure. For instance, halogenation of the acidic hydrogen atom of the imidazole ring, CR-H, would prohibit hydrogen bonding between the cation and the anion, as well as between the cation and polar solvent molecules. Hydrogen bonding is an important intermolecular interaction. Its partial or complete absence is able to drastically change physical chemical properties of the condensed phase. Such chemical modification is interesting to adjust shear viscosity of ionic liquids (e.g. for electrochemical applications) and to tune a gas capture behavior of the imidazolium cation.

The derived thermodynamics potentials, strictly speaking, correspond to an ideal gas approximation, as they are based on the molecular partition functions. Nevertheless, the observed trends must be useful in planning future functionalization of the imidazolium-based ionic liquids to develop task-specific compounds.[23-26]

**Acknowledgments**

This project is partially funded by CAPES (Brazil). Profs. Eudes Fileti and Thaciana Malaspina are hereby acknowledged for introduction to the Brazilian research community.

**Authors Information**

E-mails for correspondence: vvchaban@gmail.com. Tel: +55 12 3309-9573; Fax: +55 12 3921-8857.